\documentclass{article}
\twocolumn
\usepackage{graphicx}
\usepackage{geometry}
\usepackage{algorithm}
\usepackage{algpseudocode}
\usepackage{dirtytalk}
\usepackage{verbatim}
\usepackage{listings}
\usepackage{authblk}
\usepackage{cite}
\usepackage{url} 
\usepackage[version=3]{mhchem}
\usepackage[dvipsnames]{xcolor}
\title{Pourbaix-paper}
\author[1,*]{S. Americo}
\author[2]{I. E. Castelli}
\affil[1]{CAMD, Department of
Physics, Technical University of Denmark, DK - 2800 Kongens Lyngby,
Denmark}
\affil[2]{Department of
Energy Conversion and Storage, Technical University of Denmark, DK - 2800 Kongens Lyngby,
Denmark}
\author[1]{K.S. Thygesen}
\affil[*]{Email: stefano.americo@scitec.cnr.it}
\date{}

\begin{document}

\title{Predicting aqueous and electrochemical stability of 2D materials from extended Pourbaix analyses}
\maketitle

\begin{abstract}
A key challenge for computational discovery of electrocatalytic materials is the reliable prediction of thermodynamic stability in aqueous environment and under different electrochemical conditions. In this work, we first evaluate the electrochemical stability of more than 3000 two-dimensional (2D) materials using conventional Pourbaix diagrams (CPDs). Due to the complete neglect of thermodynamic barriers along the (often complex) reaction pathways, the vast majority of the materials are predicted to be unstable even though some are known to be stable in practice. We then introduce an analysis based on the surface Pourbaix diagram (SPD) including 'early intermediate states' that represent the first steps of the key surface passivation and dissolution reactions. The SPD framework is applied to the 2D materials MoS$_2$, phosphorene, and the MXene \ce{Ti2C}, all of which are predicted to be unstable by the CPD. For \ce{MoS2}, our approach reproduces the experimental pH-\textit{U} stability window as well as the experimental desulphurization potential. For phosphorene and \ce{Ti2C}, the SPD approach is used to investigate the spontaneous degradation mechanism and the potential-dependent surface termination, respectively, again yielding good agreement with experiments. The SPD-based stability analysis emerges as a versatile and quantitative method for prediction of stability and investigation of surface structures in electrochemical environments.
\end{abstract}

\paragraph{Introduction}

First-principles calculations are increasingly used for in-silico discovery of new materials with optimised properties for applications within a variety of areas including (electro)catalysis, photovoltaics, thermoelectrics, batteries, and nanoelectronics \cite{greeley2006computational,madsen2006automated,curtarolo2013high,kirklin2013high,ornso2013computational,zhang2019computational,chen2016understanding,hachmann2011harvard,bhattacharya2015high,castelli2012computational,hautier2013identification,yu2012identification,kuhar2018high,aykol2016high,mounet2018two,chen2015design}. 
A basic question that any such study must address, independent of the target application, is whether the predicted materials will be stable under the relevant operating conditions. The simplest and most widely used measure of thermodynamic stability is the energy above the convex hull $\Delta H_{\mathrm{hull}}$, expressing the likelihood of a material to decompose into competing solid phases in the same compositional space. Although suitable for preliminary stability assessments, this descriptor is insufficient whenever the material under investigation is meant to be employed in setups where it can exchange chemical species with the surrounding environment. Depending on the envisioned application of the material and the specific conditions under which it should operate, it will typically be necessary to consider different processes and effects beyond the solid-state reactions covered by the convex hull analysis. In most cases, degradation processes such as corrosion and dissolution due to interactions with air or moisture, will be of relevance. 
In this work we shall be particularly interested in conditions applying to electrochemical setups, where undesired charge-transfer reactions with an aqueous electrolyte aided by applied electrostatic potentials often threaten the long-term stability of electrode materials. As a special case, corresponding to zero applied potential, this covers the aqueous stability, \textit{i.e.}, the stability towards spontaneous charge transfer reactions in contact with moisture. 

The electrochemical stability issue is of paramount importance for all nanoscale structures, in particular for two-dimensional (2D) materials\cite{wang2012electronics,manzeli20172d,shanmugam2022review,ferrari2015science}. The class of 2D materials has been under intense scrutiny for about a decade. This is not only due to the unique physical properties exhibited by many of these materials but also because of their tunability, structural flexibility, and extreme surface-to-volume ratio which makes them natural candidates as battery electrodes or (electro)catalysts\cite{gogotsi2019rise,shi2017recent,pandey2017two,pandey2015two,kibsgaard2012engineering,seh2016two}. 2D materials are atomically thin and thus cannot be kinetically stabilised by \textit{e.g.}, formation of surface oxide layers as happens for many bulk materials, without significantly altering their basic properties. Methods that go beyond the 'convex hull paradigm' and allow for systematic and realistic assessment of stability under different environmental conditions, are therefore of particular relevance for this class of materials.  

Traditionally, the stability of materials in electrochemical systems is analysed by means of Pourbaix diagrams\cite{pourbaix1966atlas} introduced by Marcel Pourbaix in 1966 in his study of corrosion of metals. These potential-pH phase diagrams identify the thermodynamically stable phases of a material allowed to react with water and exchange electrons with a counter electrode. 
Conventional Pourbaix diagrams (CPD) as introduced by Pourbaix constitute a double-edged sword when it comes to material stability predictions. On one hand, due to the negligible computational cost of the method it is possible to rapidly generate pH-potential diagrams for thousands of materials. This allows, for instance, to screen large 
databases for materials that are thermodynamically stable under selected operating conditions \cite{castelli2014calculated,shinde2017discovery,back2020discovery}. On the other hand, the inherent limitations of the method, of which the most important is the complete neglect of energy barriers and reaction kinetics, can lead to erroneous conclusions regarding material stability. 
As a general rule, the more intermediate reaction steps are involved in a given process, \textit{i.e.}, the more complex the reaction is, the less likely it is to occur in practice. This is because more complex reactions involve more basic kinetic and energetic barriers separating the reactants from the products. In such situations, the system may be trapped in a metastable configuration and the entire process prevented from occurring, even if thermodynamically favorable.

In this work, we begin by performing a CPD analysis of the aqueous and electrochemical stability of more than 3000 two-dimensional (2D) materials lying on the convex hull, \textit{i.e.}, predicted as thermodynamically stable in the absence of other chemical species. This analysis shows that only a small fraction of the materials - the vast majority of which are oxides - are predicted to maintain stability under electrochemical conditions relevant for the hydrogen evolution (HER) and oxygen evolution reactions (OER). 
Surprisingly, several 2D materials that are known to be environmentally stable in practice are found to be unstable by the CPD analysis. We hypothesize that this is due to the presence of significant energy barriers associated with elementary reaction intermediates occurring along the overall reaction path. To take such barriers into account we employ an extended surface Pourbaix diagram (SPD) framework based on a concept of 'early intermediates' including relevant vacancies and adsorption configurations. A detailed stability analysis for three selected 2D materials, namely MoS$_2$, phosphorene, and \ce{Ti2C}, shows that the SPD analysis is consistent with experimental data, highlighting the necessity to go beyond the CPD for making realistic assessments of aqueous and electrochemical stability of solids in general, and for 2D materials in particular. 

We stress that the employed SPD-based analysis is completely general and can be used to analyze stability of solid surfaces in general. The methods used to calculate the conventional and surface Pourbaix diagrams are available in the Atomic Simulation Environment (ASE) package\cite{larsen2017atomic} and the calculated CPD of thermodynamically stable 2D materials are available in the C2DB database\cite{haastrup2018computational,gjerding2021recent}.

\section*{Results and discussion}

    \begin{figure*}[h!]
        \centering
        \includegraphics[width=0.85\linewidth]{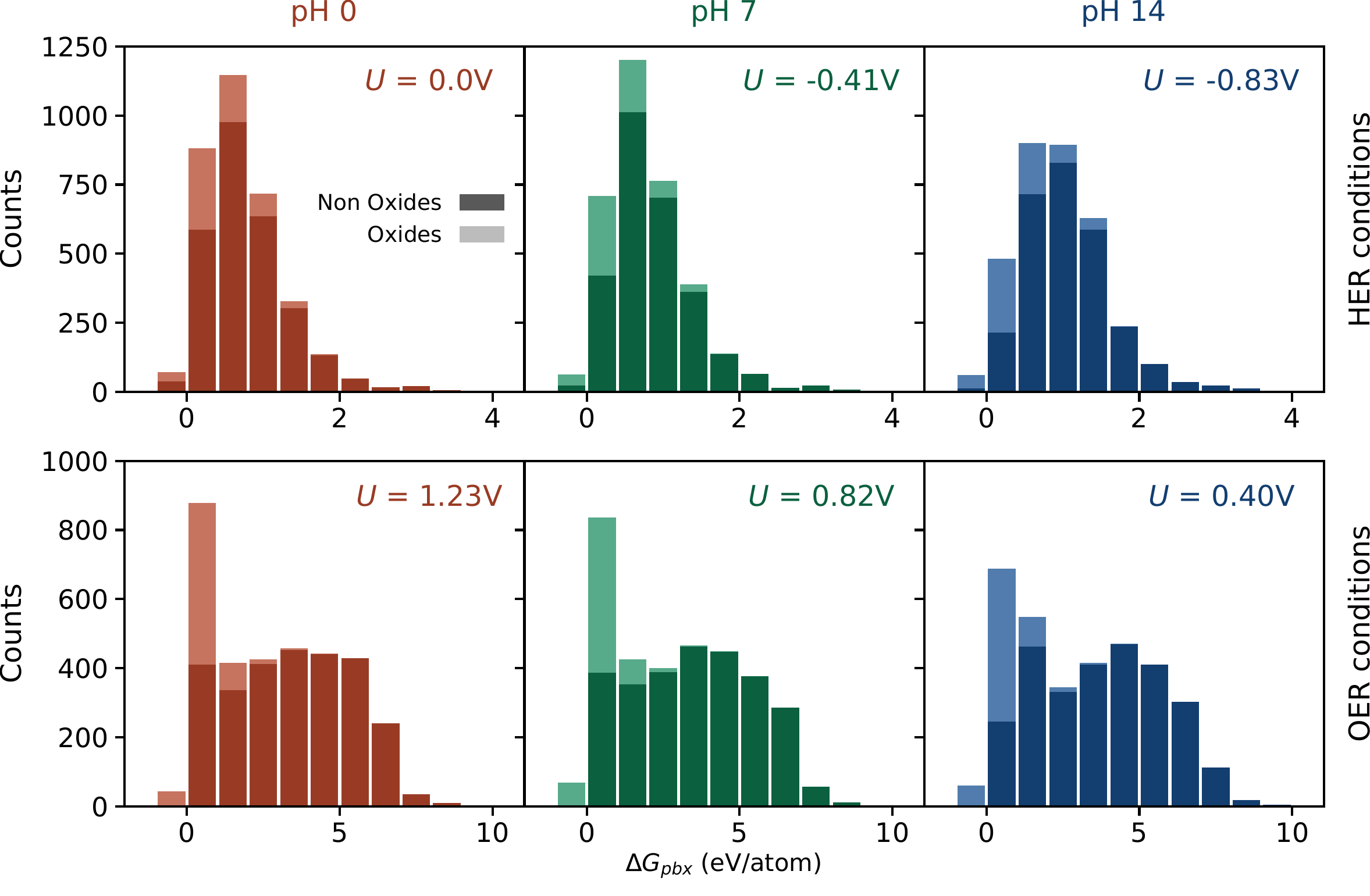}
        \caption{
                    Distribution of the Pourbaix energy $\Delta G_{pbx}$ calculated for 2902 2D materials with $E_{\mathrm{hull}}<50$ meV/atom in different conditions of pH and applied potentials vs. SHE relevant for the HER and OER processes.
                    The dataset is split between oxides (darker color) and non-oxides (brighter color).
        }
        \label{fig:pbx-stats}
    \end{figure*}

    \begin{table*}[h!]
    \renewcommand{\arraystretch}{1.5}
    \centering
        \begin{tabular}{lcccc}
        \hline
        Conditions  & \textit{U} (V) & pH & \% stable materials (of which oxides)\\ \hline
        Acidic HER   & 0          & 0  & 2.1 (46.5)      \\
        Neutral HER  & -0.41      & 7  & 1.8 (65.1)      \\
        Alkaline HER & -0.83      & 14 & 1.8 (78.7)      \\
        Acidic OER   & 1.23       & 0  & 1.3 (100.0)      \\
        Neutral OER  & 0.82       & 7  & 2.0 (100.0)      \\
        Alkaline OER & 0.40       & 14 & 1.8 (100.0)      \\ \hline
        \end{tabular}
        \vspace{10pt}
        \caption{
            Stable 2D materials in different conditions of pH and applied potential vs. SHE. The column ``\% stable materials" shows the global percentage of materials with $\Delta G_{pbx}$=0 over the total of 2902 selected materials, in each condition. Among the materials with $\Delta G_{pbx}$=0, the column ``\% stable oxides" shows, in percentage, how many are oxides. The numbers in parentheses show the percentage of stable oxides relative to the total number of oxides (604) in the same conditions}
        \label{tab:pbx-stats}
    \end{table*}

    \begin{figure*}[h!]
        \centering
        \includegraphics[width=\linewidth]{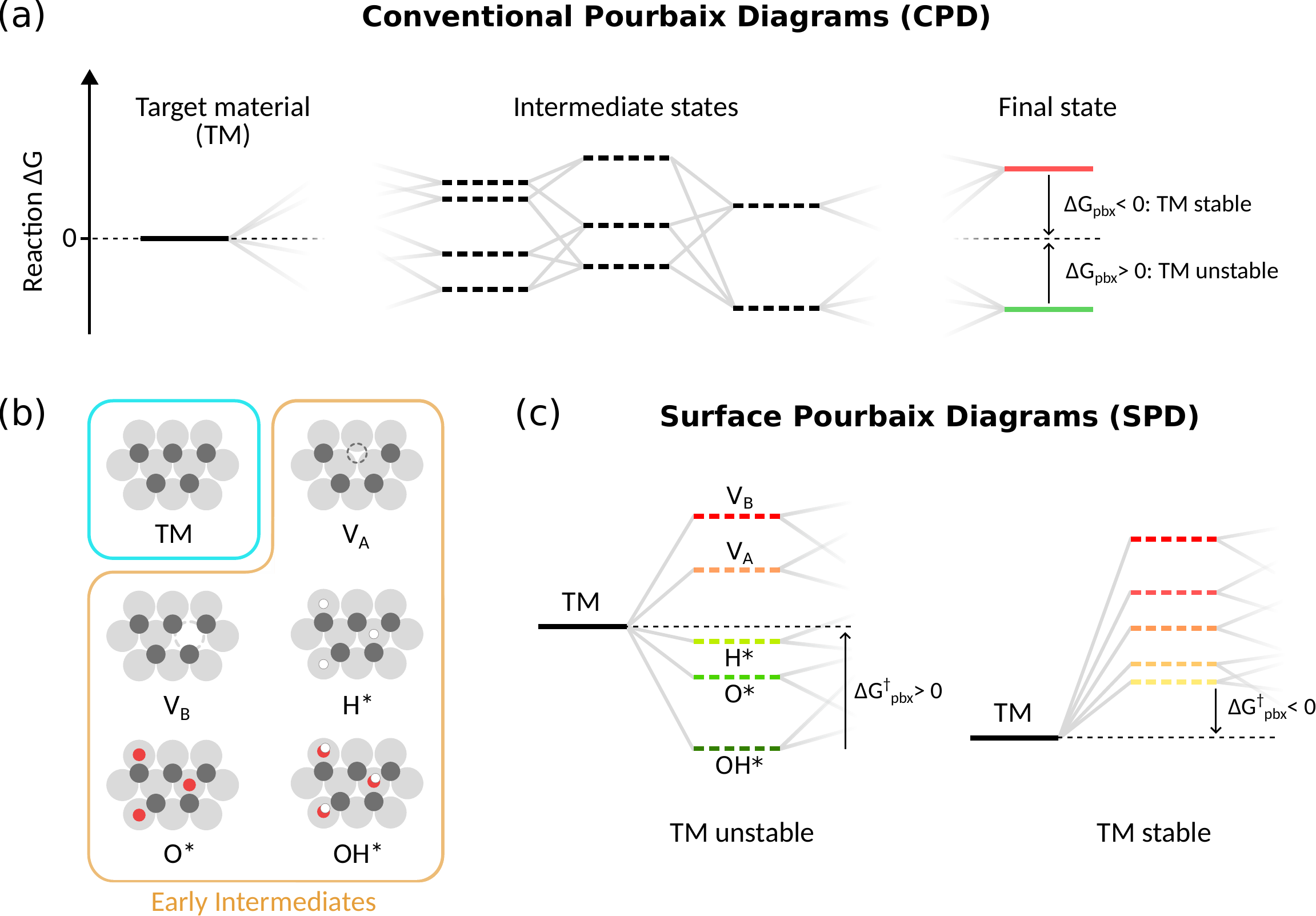}
        \vspace{5pt}
        \caption{
            a) Representative energy diagram of two different chemical reactions leading to the formation of product A and B. Product A is thermodynamically unstable with respect to the target material (black line), resulting in a negative $\Delta G_{pbx}$. However, the early reaction intermediates are downhill in free energy. The reaction path leading to product B represents the opposite scenario.
            b) Representative energy distribution of early reaction intermediates considered in the surface Pourbaix diagram construction in the case where the target material (black line) is thermodynamically unstable ($\Delta G_{pbx}>0$).
            c) Analogous to b), this time in a scenario where the target material is stable  ($\Delta G_{pbx}<0$).
        }
        \label{fig:CPD-SPD-illustration}
    \end{figure*}

\subsection*{Conventional Pourbaix analysis}
In order to obtain a general picture of the electrochemical stability of 2D materials according to the conventional Pourbaix diagram formulation, we first extracted from the C2DB database a set of thermodynamically stable monolayers as defined by $\Delta H_{\mathrm{hull}}<$50 meV/atom.
The ASE Pourbaix diagram implementation was then employed on the resulting 3376 materials in order to evaluate $\Delta G_{\mathrm{pbx}}$ (see Methods) at six relevant points corresponding to the hydrogen evolution reaction (HER) an oxygen evolution reaction (OER) under acidic, neutral and alkaline conditions, respectively.
As seen in Fig. \ref{fig:pbx-stats}, under HER condictions, the $\Delta G_{\mathrm{pbx}}$ distribution is located between 0 and 2 eV/atom. On the other hand, under OER conditions, the materials are significantly less stable and the $G_{\mathrm{pbx}}$ distribution ranges up to 8 eV/atom.
This is because at these high potentials, oxidized decomposition products are stabilized (see Eq. \ref{eq:gibbs-pourbaix}). Such products, in particular bulk metal oxides and solvated oxyanions, represent the most stable competing phases for most materials. The 2D oxides, which constitute 18\% of the entire data set, are characterized by generally higher electrochemical stability compared to all other 2D materials especially under OER conditions. In particular, all the materials with $\Delta G_{\mathrm{pbx}}<0$ are found to be oxides. In contrast, under HER conditions, the associated negative potentials promote reduction processes, which can harm the stability of 2D oxides. Still, the oxides constitute a very large fraction of all the stable materials under HER conditions as seen in Table \ref{tab:pbx-stats}. A list of all the materials found as highly stable ($\Delta G_{\mathrm{pbx}} \leq -50$ meV/atom) under at least one of the explored conditions is reported in the supplementary material, Table 1 and 2.

Overall, the number of predicted electrochemically stable 2D materials ($\Delta G_{\mathrm{pbx}}<0$) is very low and never exceeds 2.1\% of the total set of materials for any pH-voltage value. It is, however, interesting that several materials that are known to be electrochemically stable in practice, such as \ce{MoS2}\cite{li2017edge}, \ce{NbS2}\cite{yang2019ultrahigh}, graphene\cite{deng2015enhanced} and hexagonal BN\cite{ozden2019interface}, have $\Delta G_{\mathrm{pbx}}>0$ are thus predicted as unstable by the CPD analysis. 

\subsection*{Surface Pourbaix analysis}
As mentioned in the introduction, the CPD analysis neglects energy barriers along the reaction pathway and focuses on the stability of the target material relative to the final decomposition products. For that reason it gives no information on the potential \emph{kinetic} stability of the material. The inclusion of intermediate states is particularly relevant when evaluating the stability of materials containing two or more elements (other than O and H), which require to consider multiple decomposition products in order to obtain balanced chemical reactions. For instance, 
every considered reaction pathway of \ce{MoS2} has to include among the products both a Mo-containing species such as \ce{MoO3} and an S-containing species, \textit{e.g.}
\begin{equation}
    \ce{MoS2} + 7\ce{H2O} \longrightarrow \ce{MoO3} + \ce{SO4^{2-}} + 14\ce{H^+} + 12e^-
\end{equation}
Complex processes like the one above, involving the exchange of several electrons and protons as well as significant structural rearrangements leading from the starting material (\ce{MoS2}) to its main decomposition products (\ce{MoO3} and \ce{SO4^{2-}}), can be broken down into 
a large number of intermediate reaction steps, as illustrated in Fig. \ref{fig:CPD-SPD-illustration}a. By only considering the target material and its final decomposition products, the likely occurrence of reaction intermediates with unfavourable thermodynamics is neglected. 

Given that electrochemical processes take place at the interface between the electrode material and the solution phase, the early stage of any reaction will involve the atoms at the surface (for a 2D material this could be all atoms of the material depending on its thickness). 
Considering the pristine surface/2D material as the initial, "non-degraded" configuration, all degradation processes are necessarily initiated by either the adsorption of species such as O, OH and H from the solution phase or by the dissolution of atoms in the first few atomic layers towards the solution phase leaving a surface vacancy\cite{rong2015ab}.
We refer to such adsorption and vacancy configurations - illustrated in Fig. \ref{fig:CPD-SPD-illustration}b - as "early intermediate states", and to the processes leading from the pristine surface to one of these configurations as "early degradation steps" (EDS). The Gibbs free energy change of the EDS is labeled as $\Delta G^{\dag}_{pbx}$.
The EDS can be energetically costly, especially without the aid of large applied potentials and/or harsh pH conditions, leading to steep thermodynamic energy barriers that prevents the global degradation process to take place altogether. The energy diagram on the left in Fig. \ref{fig:CPD-SPD-illustration}c illustrates a scenario where one or more EDSs are thermodynamically favourable, hence the target material is labeled as unstable (positive $\Delta G^{\dag}_{pbx}$). Conversely, if all the EDS are uphill in energy as in the case reported on the right, the material is considered as meta-stable (negative $\Delta G^{\dag}_{pbx}$). Note that with this definition a meta-stable material could be globally stable (if $\Delta G_{pbx}<0$).

In order to include these early intermediates we construct surface Pourbaix diagrams (SPD) according to the procedure described in the Methods section. Our formulation also includes a correction scheme - described in detail in the supplemental material - to account for the surface excess or depletion of ions at the charged metal-solution interface.
In the following, we apply the SPD framework to three selected 2D materials, namely \ce{MoS2}, phosphorene and \ce{Ti2C}. Comparisons with the CPD and experimental data highlight the great flexibility and quantitative accuracy of the SPD-based analysis for describing the behaviour of real electrochemical systems.

\paragraph{\ce{MoS2}}

\begin{figure*}[ht!]
       \centering
            \vspace{20pt}
            \includegraphics[width=\linewidth]{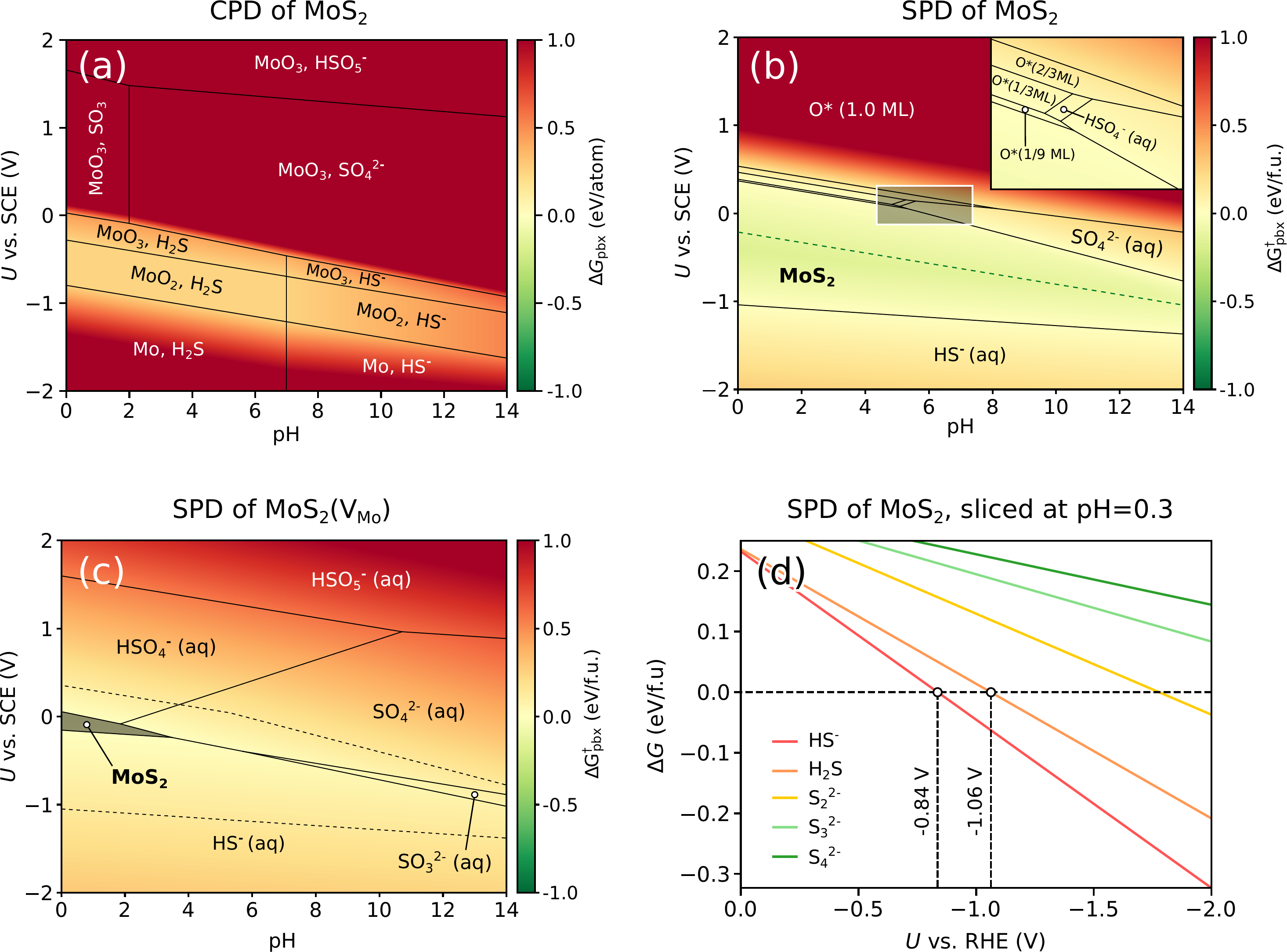}
            \caption{
                a) Conventional Pourbaix diagram of \ce{MoS2}, with SCE as the counter-electrode. The color map shows the Pourbaix energy per atom.
                b) Surface Pourbaix diagram of \ce{MoS2} with SCE as the counter-electrode. The color map shows the Pourbaix energy per unit formula. An inset covering the darkened region is shown at the top right in order to better highlight the phases associated with narrow stability domains. The dashed green line represents the HER equilibrium vs. SCE. The concentration of \ce{SO_4^{2-}} ions has been set to the experimental bulk value of 1.0 mol l$^{-1}$ and then corrected for the surface excess (see supplemental material) in order to compare with the CV data in refs. \citen{MoS2-supercap-1} and \citen{MoS2-supercap-2}.
                c) Surface pourbaix diagram obtained from a \ce{MoS2} surface with a Mo vacancy. The darkened region represent the new, reduced stability window of \ce{MoS2} due to the stabilization of S vacancies around the Mo defect. The dashed lines show the stability window of pristine \ce{MoS2}, as in panel b.
                d) Slice at pH=0.3 of the SPD of \ce{MoS2} vs. RHE. The colored lines describe the energy profile of the dissolution EDS producing sulphur-containing species.
                The vertical dashed line at -0.84 V marks the predicted onset of the overall desulphurization process. The vertical line at -1.06 V marks the onset of \ce{H2S} evolution.
            }
            \label{fig:MoS2-panel}
    \end{figure*}

Molybdenum disulphide \ce{MoS2} is one of the most extensively studied materials in the class of transition metal dichalcogenides (TMDs) due to its unique properties and potential applications in various fields, including electronics, optoelectronics, catalysis, and more\cite{MoS2-applications,choi2014lateral, lukowski2013enhanced}. A main reasons for its success in several fields of application is its good stability under a wide range of experimental conditions. In particular, the use of \ce{MoS2} as catalyst for the HER\cite{MoS2_HER_applications} and for water treatment\cite{MoS2_water_treatment} highlights the stability of the material in aqueous environments in a wide range of pH conditions and, at least, moderate potentials required for the HER.

Existing studies on \ce{MoS2} for supercapacitor applications\cite{MoS2-supercap-1,MoS2-supercap-2} provide useful cyclic voltammetry (CV) profiles at pH 0 and applied potentials ranging between +0.5 V and -1.1 V against the saturated calomel electrode (SCE). No significant oxidation or reduction peaks are found within this potential window, confirming the material stability. As shown in Fig. \ref{fig:MoS2-panel}a, the CPD generated herein for \ce{MoS2} predicts the material to be thermodynamically unstable across all pH conditions and potentials, in contradiction with the experimental observations. This is due to the low formation energies of molybdenum oxides such as \ce{MoO3} and \ce{MoO2}, as well as sulfur oxyanions and molecular compounds such as \ce{SO4^{2-}} and \ce{H2S}.

Note that all the \ce{MoS2} decomposition pathways defining the stability domains in Fig. \ref{fig:MoS2-panel}a (such as the oxidation to \ce{MoO3} and parallel dissolution into \ce{SO4^{2-}}) imply complex transformations of the initial structure, and are poorly described by a single thermodynamic step.
Shifting our focus to the EDSs instead, we generate the SPD shown in Fig. \ref{fig:MoS2-panel}b.
A large stability domain is now found, spanning both positive and negative potentials around the HER equilibrium potential (dashed green line) across the entire pH range.
The newfound stability of \ce{MoS2} towards dissolution processes is due to the positive and high vacancy formation energies which, at a vacancy density of 0.012 \AA$^{-2}$ (one vacancy every nine unit cells) and relative to the vacant elements in their standard states, amounts to 2.97 eV (S vacancies) and 7.74 eV (Mo vacancies). The stability towards surface passivation is due to \ce{MoS2} the low basal plane reactivity\cite{MoS2basalplanesucks-1,MoS2basalplanesucks-2,MoS2basalplanesucks-3} leading to positive adsorption energies of O, OH and H adsorbates (0.72, 1.80 and 2.20 eV, respectively).

The SPD diagram in Fig. \ref{fig:MoS2-panel}b shows excellent agreement with the aforementioned CV scans, identifying a stability window at pH 0 between +0.3 and -1.1 V. Above 0.3 V, surface passivation with atomic oxygen is predicted to be favorable. In comparison, the experiments find the material to be stable up to at least 0.5 V. Nevertheless, both cited papers\cite{MoS2-supercap-1,MoS2-supercap-2}, as well as a study by Chia \textit{et al.}\cite{chia2015catalytic}, report the onset of an anodic peak already at 0.3 V, suggesting the occurrence of a reversible oxidative process already in the 0.3-0.5 V region. By comparison with Fig. \ref{fig:MoS2-panel}b, this process can be attributed to the adsorption of low coverage atomic oxygen.

The diagram in Fig. \ref{fig:MoS2-panel}b discussed so far describes early steps of degradation processes affecting a pristine \ce{MoS2} reference structure.
In principle, however, the pristine structure is an idealization and any real material will contain defects, e.g, vacancies or other types of point defects, even if they are thermodynamically unstable. Such defects may act as nucleation centers for extended defects such as vacancy pairs, clusters, or line defects, which could eventually lead to the degradation of the material. This means that the pristine crystal structure/surface may not be a realistic reference structure for the SPD analysis. In Fig. \ref{fig:MoS2-panel}c, we investigate the role of native defects on the stability of \ce{MoS2} towards dissolution by using a surface with a pre-existing Mo vacancy as the reference state and defining dissolution processes that produce a second neighboring Mo or S vacancy. We find that Mo defects greatly stabilize nearby S vacancies, whose formation energy is decreased from 2.97 eV on the pristine surface to 0.91 eV. As a consequence, the dissolution into sulfur-containing ions and molecules becomes accessible at more moderate potentials, significantly reducing in size the stability domain of \ce{MoS2} and making the presence of Mo vacancies in the as-synthesized material undesirable. With that said, we highlight that given their really high formation energy, a large concentration of Mo vacancies is unlikely to occur experimentally.
The same cooperative effect is not seen for other combinations of defect pairs. For instance, forming an S vacancy next to a pre-existing S vacancy yields practically the same formation energy (2.96 eV) as creating the first S vacancy on the pristine surface. 
Our results suggest that the nucleation of extended defects on \ce{MoS2} is unlikely: after the S atoms around a pre-existing Mo vacancy are preferentially dissolved, subsequent dissolution steps do not gain a significant advantage from the newly formed S vacancies and an eventual radial expansion of the defect is prevented.

Finally, a study by Tsai \textit{et al.} investigates the effect of electrochemically induced sulphur vacancies at the basal plane of \ce{MoS2} on the HER catalytic activity\cite{tsai2017electrochemical}. The experimental onset of the desulphurization process is observed around -0.5 V vs. RHE. The S vacancy concentration reaches an equilibrium value at about -1.0 V, where \ce{H2S} formation is predicted theoretically to be favourable. These observations are in good agreement with the results presented in fig. \ref{fig:MoS2-panel}d, showing the energy profile as a function of the electrode potential of different dissolution processes involving the formation of S vacancies, after imposing the experimental pH value of 0.3. At -0.84 V vs. RHE, \ce{MoS2} dissolves favourably into \ce{HS-} ions, which can be related to the experimental onset of the desulphurization process. At -1.06 V \ce{H2S} formation becomes favourable as well, in perfect agreement with the theoretical results in the cited work.

\paragraph{Phosphorene}

    \begin{figure*}[ht!]
       \centering     
            \vspace{20pt}
            \includegraphics[width=\textwidth]{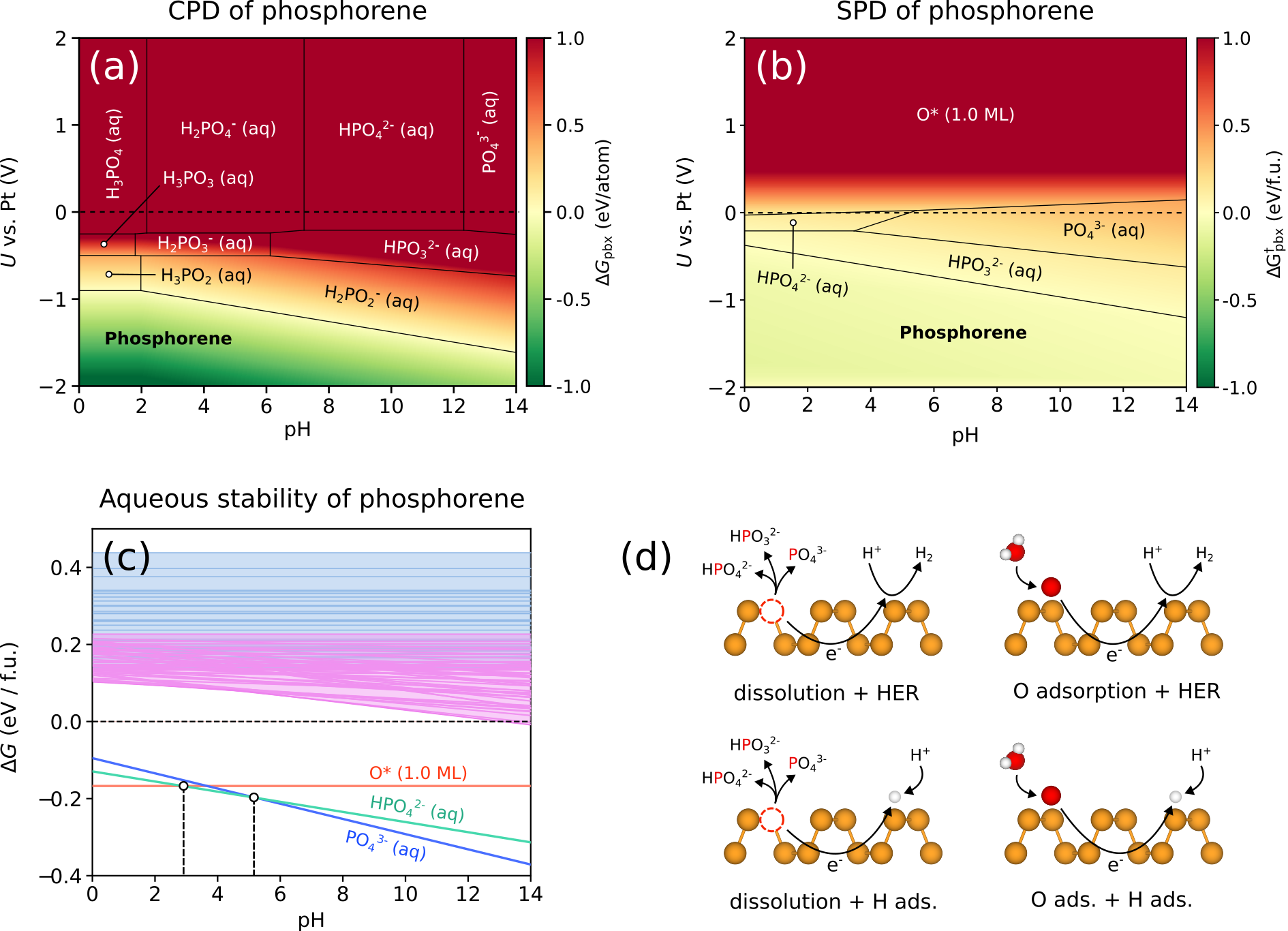}
            \caption{
                a) Conventional Pourbaix diagram of phosphorene, with Pt as the reference electrode.
                b) Surface Pourbaix diagram of phosphorene, with Pt as the reference electrode. The dashed black line at U=0V vs. Pt highlights the relevant region for stability predictions about spontaneous processes.
                c) Slice of the diagram in panel b along the dashed line at U=0V, obtained by including mixed surface states. The light blue lines forming a band at positive free energies represent all the mixed adsorption-adsorption states. The pink band contains all the mixed adsorption-dissolution states. The three colored lines at negative free energies represent the free energy profile of the most stable phases found along the 0 V line in panel b.
                Vertical dashed lines highlight the pH corresponding to the the phase boundaries.
                d) Atomic models illustrating the four different types of processes present in panel (c).
            }
            \label{fig:phosphorene-panel}
    \end{figure*}
    
Phosphorene is the 2D counterpart of black phosphorus, which has a layered structure in bulk phase and can be mechanically exfoliated. Due to its high hole mobility and tunable band gap, phosphorene has been identified as a promising candidate for applications in field-effect transistors and  photodetectors\cite{liu2014phosphorene,carvalho2016phosphorene,eswaraiah2016black}.
One of the main practical limitations towards the use of phosphorene in opto-electronic devices is its poor stability in presence of oxygen and moisture, making its encapsulation in more stable materials a necessary precaution\cite{sang2019recent,ziletti2015oxygen,rahman20162d}.  

As expected, the CPD in Fig. \ref{fig:phosphorene-panel}a predicts phosphorene to be unstable under ambient conditions, represented by the dashed line at $U$=0 V. This is mainly due to the wide range of stable oxidation states of phosphor and, accordingly, of solvated oxyanions \ce{H_{x}P_{y}O_{n}^{z-}} with low formation energies.
From a microscopic point of view, the phosphorene surface is rather reactive: oxygen adsorption is exothermic under standard conditions (-56.1 meV at 25\% coverage) and the P vacancy formation energy is only 1.70 eV. Accordingly, some of the available dissolution and passivation pathways have negative $\Delta G$ in absence of an applied potential and, similarly to the CPD, the SPD in Fig. \ref{fig:phosphorene-panel}b predicts the material to be unstable in ambient conditions.

As observed by Wood \textit{et al.}, phosphorene samples in ambient conditions degrade in few days. As the process is monitored, an XPS peak attributed to oxidized phosphor species ($\ce{PO_x}$) appears with concurrent formation of bubbles on the sample surface\cite{wood2014effective}. These are clear signs of a spontaneous reaction resulting in the oxidation of the surface P atoms and the concurrent evolution of a gaseous species. 
Fig. \ref{fig:phosphorene-panel}c shows the energy profile of different spontaneous degradation pathways of phosphorene in ambient conditions, described by the \textit{U}=0 dashed line in fig. \ref{fig:phosphorene-panel}b. The EDS found along the latter (1.0 ML O passivation, dissolution into \ce{HPO4^{2-}}/\ce{PO4^{3-}}) - as well as several other dissolution and passivation processes not shown here - are all thermodynamically favorable over the whole pH range and can take place simultaneously without the aid of an applied potential. These oxidative processes are paralleled by the catalytic conversion of solvated protons into \ce{H2}, as shown by the atomic models at the top of Fig. \ref{fig:phosphorene-panel}d. Mixed surface processes (see the two models at the bottom of Fig. \ref{fig:phosphorene-panel}d) form the pink band (H adsorption + dissolution) and blue band (O adsorption + H adsorption) in Fig. \ref{fig:phosphorene-panel}c. Since these are mildly endoergonic over the entire pH range, they are not expected to contribute to the material degradation.
Our analysis suggests that, after a prolonged exposure to the atmospheric moisture, phosphorene releases a variety of phosphor oxoanions while adsorbing oxygen, leaving a partially dissolved, partially passivated surface. The XPS peaks observed in the cited study can be attributed to oxoanions such as \ce{HPO4^{2-}} and \ce{PO4^{3-}}, which can remain adsorbed on the surface, as well as the P atoms in contact with the adsorbed oxygen. All of these degradation processes are paralleled by hydrogen evolution, the most logical origin of the bubbles observed experimentally.

\paragraph{\ce{Ti2C} MXene}

    \begin{figure*}[ht!]
       \centering     
            \includegraphics[width=\textwidth]{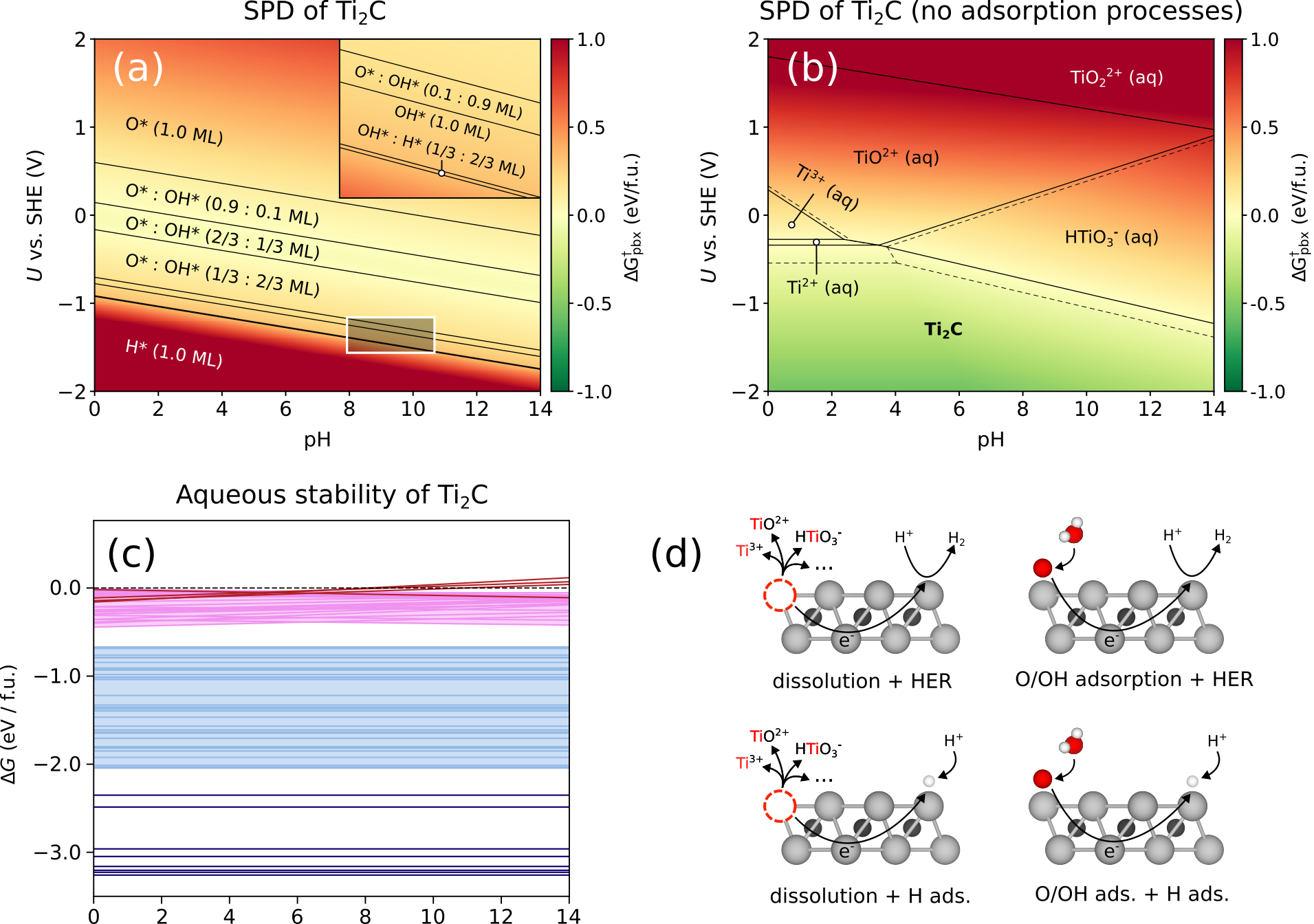}
            \caption{
                a) Surface Pourbaix diagram of \ce{Ti2C} with SHE as the reference electrode, including mixed adsorpion states. The color map shows the Pourbaix energy per unit formula. For better visualization, the reference system chosen for defining the pourbaix energy scale is the 2/3ML O*, 1/3ML OH* mixed adsorption state. An inset covering the darkened region is shown at the top right in order to better highlight the phases associated with narrow stability domains.
                b) Surface pourbaix diagram of \ce{Ti2C} obtained by considering only the dissolution processes, \textit{i.e.}, adsorption configurations are omitted during the diagram construction.
                An analogous diagram, obtained by using the surface with a pre-exisitng Ti vacancy as the reference state, is overlaid (dashed lines).
                c) Slice of the diagram in panel (a) along the dashed line at U=0V. Mixed surface reactions are included. The light blue lines forming a band at positive free energies represent all the mixed adsorption reactions. The pink band contains all the mixed adsorption-dissolution reactions. The dark blue lines represent adsorption processes paralleled by hydrogen evolution. The three dark red lines represent the most stable dissolved phases of \ce{Ti2C} (\ce{Ti^{3+}}, \ce{TiO^{2+}}, \ce{HTIO3-}), as found in panel (b) along the zero line.
                d) Atomic models illustrating the four different types of processes present in panel (c).
                }
                \label{fig:surfpbx-Ti2C}
    \end{figure*}

MXenes are a class of 2D materials that belong to the family of transition metal carbides, nitrides, and carbonitrides (M: transition metal; X: carbon and/or nitrogen). Often characterized by exceptional electronic, mechanical, and thermal properties, these materials are highly versatile for a range of applications including batteries\cite{ming2021mxenes}, gas sensing\cite{lee2017room} and photocatalysis\cite{zhong2021two}.
Titanium carbide \ce{Ti_xC_y} is among the most studied MXenes. It is characterized by a rich surface chemistry which poses challenges towards its use in electrochemical applications.
For instance, \ce{Ti_xC_y} electrodes have been recently tested as HER catalysts, showing only modest performance which tends to further decrease over several cycles\cite{seh2016two}.
As observed by Zhang \textit{et al.}, colloidal suspensions of \ce{Ti3C2} can decompose to \ce{TiO2} and amorphous carbon over several days in solution\cite{zhang2017oxidation}. It has also been found that the material tends to form passivation layers \cite{fredrickson2016effects,gao20172d} and to decompose in aqueous environments.

The general instability of \ce{Ti_xC_y} in an aqueous environment together with its rich surface chemistry makes this system a perfect case study for our SPD framework. We choose as our model surface the thinnest possible form of titanium carbide, namely \ce{Ti2C}. The CPD of \ce{Ti2C} shown in Fig. S1 predicts the material to be unstable in agreement with the experimental evidence, but provides no information about its surface state. In order to cover a wider range of surface terminations, we model the adsorption of O, OH and H at various coverages (0.1ML, 0.33ML, 0.66ML, 1.0ML) and we include mixed coverage states.
The SPD, shown in Fig. \ref{fig:surfpbx-Ti2C}a, highlights the remarkable surface reactivity of \ce{Ti2C}, which favors the coverage by a full monolayer of adsorbates in the whole spanned $U$-pH region. As expected, the overall oxidation state of the adsorbate layer progressively increases when moving from negative to positive applied potentials, going from a ML of adsorbed hydrogen up to a ML of oxygen. Mixed coverage states are found stable at intermediate potentials.
These results agree well with the surface Pourbaix diagram obtained by Gao \textit{et al.}\cite{gao20172d}, with the difference that in our SPD we include hydrogen adsorption, which becomes stable at negative potentials.

We note that surface vacancies are not found to be energetically favorable under any voltage or pH conditions. Nevertheless, several of the dissolution processes have negative $\Delta G$ in extended regions of the diagram, as seen in the SPD in Fig. \ref{fig:surfpbx-Ti2C}b, obtained by masking all the adsorption configurations and thus showing only vacancy EDS. Similar to the case of \ce{MoS2}, pre-existing Ti vacancies are found to stabilize neighboring ones, although to a smaller extent: their formation is reduced from 2.89 eV on the pristine surface to 2.34 eV on a defective one. This causes some of the dissolved phases (especially \ce{Ti^{2+}} and \ce{HTiO3^-} to extend their stability domain by a small, but significant amount (see the dashed lines). 

In order to investigate spontaneous degradation processes, we report in Fig. \ref{fig:surfpbx-Ti2C}c a slice of the SPD against the Pt electrode at $U=0$. Despite passivation processes being the most favourable ones from a thermodynamic standpoint, at least one of the "pure" dissolution EDS (red lines) is exoergonic in any pH conditions, supporting the observations by Xia \textit{et al.} regarding the role of Ti vacancies in \ce{Ti2C} degradation \cite{xia2019ambient}. Furthermore, all the mixed adsorption-dissolution reactions (pink band) are spontaneous across the entire pH range. This suggests a cooperative role of passivation and Ti dissolution in compromising the material stability, compatibly with the observations by Ibragimova \textit{et al.}\cite{ibragimova2022native}.

Overall, these results confirm the natural tendency of bare titanium carbide surfaces to form stable passivation layers, whose composition depends on the applied potential and pH conditions.
The dissolution of Ti surface atoms, enhanced in presence of native Ti vacancies, is confirmed to contribute to the material degradation.

\section*{Conclusions}
In this work we employ both a conventional Pourbaix diagram (CPD) and a surface Pourbaix diagram (SPD) methodology in order to investigate the electrochemical stability of solid state materials, with a focus on two-dimensional (2D) materials. Our updated CPD description implemented in the ASE python package, used as a backbone for the SPD implementation, is proven suitable for rapidly finding qualitative trends on large material datasets. With that said, we find the CPD to severely underestimate the electrochemical stability in general, as highlighted by the very low number of 2D materials found stable in HER and OER conditions. We attribute this to the neglect of early degradation steps (EDS) with unfavorable thermodynamics that can prevent degradation processes to take place. 

The EDS are captured by our comprehensive SPD framework, employed on \ce{MoS2}, phosphorene and \ce{Ti2C} as three separate case studies. The SPD grants mechanistic insight, great flexibility and predictive power when comparing to experimental results, at the expense of a higher computational cost. Using the SPD, we quantitatively reproduce the experimental desulphurization potential and stability window of \ce{MoS2}, incorrectly described by the CPD. Our results also suggest that \ce{MoS2} samples with a significant concentration of Mo vacancies are more prone to dissolution compared to samples with low defectivity. As for phosphorene and \ce{Ti2C}, the SPD provides additional mechanistic insight on their spontaneous degradation and surface state as a function of the pH and applied potential, supporting experimental and theoretical results in good agreement.

We stress that including vacancy configurations in the SPD is fundamental in order to cover dissolution processes, thus obtaining a complete picture of the electrochemical stability of materials in general. The accuracy of SPD results can be improved by operating on the \textit{ab initio} modeling of the surface configurations. For instance, including solvation effects or investigating different vacancy concentrations can further reduce the gap between theoretical predictions and experiments. Overall, we discourage the use of CPD results as conclusive data on the electrochemical stability of materials. We recommend instead the SPD whenever a detailed comprehension of electrochemical degradation processes is required. 

As a final remark, we point out that performance and stability are tightly connected.
With the focus in the materials science literature often being on materials properties rather than on stability, obtaining accurate models for the latter is complicated by the lack of reliable data. Advancing our understanding of (electrochemical) stability of materials can be only reached only by combining high quality experimental data with accurate computational modeling. We believe that more efforts have to be dedicated in both the experimental and computational communities to understanding fundamental degradation mechanisms and elucidating the surface structure and stability of materials.

\section*{Methods}

\paragraph{Conventional Pourbaix diagrams}
An efficient utility to calculate conventional Pourbaix diagrams was implemented in the Atomic Simulation Environment (ASE) package\cite{larsen2017atomic}.
The method initially defines a list of electrochemical reactions describing every possible decomposition pathway of the target material into competing solid and solvated phases. We retrieve the relevant solid phases from the OQMD database\cite{OQMD}, and the solvated phases from the ASE phase diagram utility.
The reaction free energy $\Delta G$ of each electrochemical process at temperature \textit{T} is determined by writing the Gibbs free energy in terms of the pH of the solution and the potential \textit{U} applied between the working electrode (represented by the target material) and a reference electrode of choice:
    \begin{equation}
        \Delta G = \Delta G^0 + k_BT\ln Q' -  n_\mathrm{H}\,\alpha~\mathrm{pH} - n_e\,U~
    \label{eq:gibbs-pourbaix}
    \end{equation}
With $\alpha = k_B T \ln{10}$. $n_\mathrm{H}$ and $n_e$ are the number of protons and electrons, respectively, exchanged in the semi-reaction at the working electrode. Their values are positive for reactants and negative for products. Electrons are among the products in oxidation processes, and among the reactants in reduction processes.

The standard reaction free energy is written as
    \begin{equation}
        \Delta G^0 = \sum^{\mathrm{react}}_r \mu^0_r n_r - \sum^{\mathrm{prod}}_p \mu^0_p n_p
        \label{eq:deltaG0}
    \end{equation}
where $n_r$ and $n_p$ denote the stoichiometric coefficients of reactants and products, respectively. The standard chemical potentials $\mu^0$ of all the solid species are represented by their formation energies with respect to the elements in their standard state. For species containing hydrogen and/or oxygen, we correct the standard state energies following Persson \textit{et al.}\cite{persson2012prediction}.
The $\mu_0$ of solvated species are obtained from thermodynamic tables\cite{pourbaix1966atlas,johnson1992supcrt92}. $Q'=\prod_{i \neq \ce{H^+}} a_i^{n_i}$ gathers the activities of the ionic species (except for H$^+$) as determined by their concentration which we set to the fixed value of $10^{-6}$ mol L$^{-1}$. The contribution from protons is considered separately in the pH term.

The code supports the use of five different reference electrodes, which are modeled by applying a zero-point shift of $-n_e U^0_{\mathrm{ref}}$ to eq. \ref{eq:gibbs-pourbaix} and including an additional contribution to the pH term when needed. The Standard Hydrogen Electrode (SHE) sets the zero of the potential scale, hence $U^0_{\mathrm{ref}}$=0V, and carries no pH dependence.
For the Ag/AgCl and Saturated Calomel Electrode (SCE) $U^0_{\mathrm{ref}}$ amounts to 0.222V and 0.244V, respectively, with no pH dependence.
For the Reversible Hydrogen Electrode (RHE), $U^0_{\mathrm{ref}}$=0V and a pH-dependent contribution of $n_e\,\alpha\,pH$ is included in eq. \ref{eq:gibbs-pourbaix}. We include the Pt electrode as well, which conceptually represents any inert metallic electrode able to perform the hydrogen evolution reaction (HER) and the oxygen evolution reaction (OER) catalytically. For all oxidation processes ($n_e>0$), the Pt electrode behaves exactly as the RHE, hence the same $U^0_{\mathrm{ref}}$ and pH correction apply. However, since a Pt electrode cannot reverse the HER or the OER as it is not supplied by \ce{H2} or \ce{O2}, all the reduction processes ($n_e<0$) are raised in energy by $n_e\,U^0_{\ce{O2}/\ce{H2O}}$, where $U^0_{\ce{O2}/\ce{H2O}}=1.23$V is the standard electrode potential of the oxygen evolution reaction (OER) - and maintain the pH dependence.
The potential profile as a function of the pH of the implemented reference electrodes is shown in Fig. S2.

When generating Pourbaix diagrams, an expression of the form of Eq. \ref{eq:gibbs-pourbaix} is pre-computed for each of the considered reaction pathways, and then evaluated as a function of the pH and applied potential \textit{U}. This allows to simultaneously determine the energy $\Delta G_{pbx}$ of the target material relative to the most stable competing phase and the identity of the latter:
    \begin{equation}
        \Delta G_{\mathrm{pbx}}(\mathrm{pH}, U) = -\min_{\mathrm{react.}} \Delta G(\mathrm{pH}, U)
    \label{eq:pourbaix-energy}
    \end{equation}
where all considered reaction pathways are included on the right hand side.
If $\Delta G_{pbx}$ is negative in a given pH-\textit{U} window, the target material is labeled as stable.

\paragraph{Surface Pourbaix diagrams}
The same ASE framework was employed (partially) to generate surface Pourbaix diagrams. Density functional theory as implemented in GPAW\cite{mortensen2024gpaw} is used to obtain the relaxed geometry and ground state energy of the reference surface and early intermediate states. We employ the PBE exchange-correlation functional\cite{perdew1996generalized} and include Grimme D3 corrections\cite{grimme2010consistent} in order to capture van der Waals interactions. A 3x3x1 supercell is utilized for all surface configurations except when considering pre-existing vacancies where 4x4x1 supercells are employed. The chemical potentials of the pristine surface and early intermediates are represented by their formation energies obtained from the D3-corrected total energies. The early degradation steps represent all the considered electrochemical processes. These are laid out by combining the reference surface configuration with each of the early intermediates and balancing out with \ce{H2O}, \ce{H^+}, electrons, and ionic species. \textit{e.g.}:
    \begin{equation}
        \mathrm{\ce{MoS2} + \ce{H2O} \longrightarrow \ce{MoS2}(OH) + \ce{H^+} + e^-}
    \end{equation}
    \begin{equation}
        \mathrm{\ce{MoS2} + 4\ce{H2O} \longrightarrow \ce{MoS2}(V_S) + \ce{SO4^{2-}} + 8\ce{H^+}+ 6e^-}
    \end{equation}
In the above reactions \ce{MoS2}(OH) and \ce{MoS2}($\mathrm{V_S}$) represent \ce{MoS2} with an adsorbed OH and with a S vacancy, respectively.
The construction of the surface Pourbaix diagram proceeds analogous to the one described for conventional Pourbaix diagrams. Pourbaix energies are normalized with respect to the number of formula units in the supercell.
A comprehensive view of all the EDS considered by the SPD framework is shown in the supplemental material, Fig. S1.

Since the modeled electrochemical processes take place in proximity of a charged surface, it is a crude approximation to set the concentration of all charged species to the same arbitrary value as routinely done in the CPD analysis. We employ a simple correction scheme in our SPD formulation, to account for the surface excess or depletion of ionic species as a function of the applied electrode potential. The theoretical background behind the implementation of the correction scheme is described in the Supporting Material.

\paragraph{Aqueous stability in ambient conditions}
When a material is in contact with an aqueous solution
without an external circuit driving electrons from or towards a counter electrode, spontaneous electrochemical degradation may still take place. In this case the material acts as both the cathode and the anode, hence the global electron transfer reaction takes place at one electrode-solution interface. Since the electrons involved in oxidation and reduction are taken/delivered at the same potential (the Fermi level of the material surface), the $n_eU$ contribution to the reaction energetics in eq. \ref{eq:gibbs-pourbaix} is zero. On the other hand, the pH of the solution still influences the reaction energetics.
Three scenarios are possible in such conditions: (i) The electrode material gets oxidized, e.g. by adsorbing oxygen, releasing electrons into another region of the surface where \ce{H2} is catalytically evolved from \ce{H^+}. (ii) The electrode material gets reduced, e.g. by adsorbing hydrogen with the necessary electrons obtained from the catalytic oxidation of \ce{H2O} into \ce{O2} in a nearby surface region. (iii) Both the oxidation and reduction semi-reactions affect the surface structure of the material through adsorption or dissolution.
Scenarios i) and ii) are practically reproduced by slicing along the $U=0$ line the SPD of the target material against the Pt electrode.
In order to reproduce scenario iii), we include in the same slice a set of reactions obtained by combining oxidation and reduction EDS's with each other, \textit{e.g}:
    \begin{equation}
        \mathrm{A) ~ \ce{MoS2} + \ce{H2O} \longrightarrow \ce{MoS2}(OH) + \ce{H^+} + e^-}
    \end{equation}
    \begin{equation}
        \mathrm{B) ~ \ce{MoS2}  + \ce{H^+} + e^- \longrightarrow \ce{MoS2}(H)}
    \end{equation}
    \begin{equation}
        \mathrm{A+B) ~ 2\ce{MoS2}  + \ce{H2O} \longrightarrow \ce{MoS2}(H) + \ce{MoS2}(OH)}
    \end{equation}

\section{Supporting Information}
List of the 72 materials found as highly stable under the CPD, CPD of the \ce{Ti2C} MXene, Implemented reference electrodes and their pH-dependent potentials, illustrations of electrochemical processes modeled by the SPD, surface excess correction scheme (PDF).

The complete list of 3376 materials used for the CPD analysis (Fig. \ref{fig:pbx-stats} and Table \ref{tab:pbx-stats}), as well as the code used for generating the SPD and acqueous stability diagrams are available at \url{https://github.com/surfpbx/surfpbx}.

\section{Acknowledgements}
S.A. and K.S.T. acknowledge funding from the European Research Council (ERC) under the European Union’s Horizon 2020 research and innovation program Grant No. 773122 (LIMA) and the Villum Investigator Grant No. 37789 supported by VILLUM FONDEN. I.E.C. acknowledge support from the Independent Research Fund Denmark (Research Project 1, project “Rational Design of High-Entropy Oxides for Protonic Ceramic Fuel Cells (HERCULES)”, grant No 1032-00269B).

\bibliographystyle{unsrt}
\clearpage
\bibliography{main.bib}
\end{document}


\title{Predicting aqueous and electrochemical stability of 2D materials from an extended Pourbaix analysis \vspace{0.5cm} \newline\LARGE{Supporting information}}
\author[1,*]{Stefano Americo}
\author[2]{Ivano Eligio Castelli}
\author[1]{Kristian Sommer Thygesen}
\affil[1]{Computational Atomic-scale Materials Design (CAMD), Department of Physics, Technical University of Denmark, 2800 Kgs. Lyngby Denmark}
\affil[2]{Department of
Energy Conversion and Storage, Technical University of Denmark, DK - 2800 Kongens Lyngby,
Denmark}
\affil[*]{Email: stefano.americo@scitec.cnr.it}
\date{}

\maketitle

\clearpage
\section*{Stable 2D materials under the CPD analysis}
\begin{table}[h!]
    \centering
    \begin{tabular}{lc|cccccc}
\hline
\multirow{3}{*}{Formula} & \multirow{3}{*}{$\Delta H_{hull}$ (eV/atom)} & \multicolumn{6}{c}{$\Delta G_{pbx}$ (eV/atom)}                                 \\ \cline{3-8} 
                         &                                              & \multicolumn{3}{c|}{HER Conditions}      & \multicolumn{3}{c}{OER conditions} \\
                         &                                              & pH 0 & pH 7 & \multicolumn{1}{c|}{pH 14} & pH 0      & pH 7      & pH 14    \\ 
			\hline
\ce{Ag2BaO8} & 0.050 & 1.437 & 1.36 & \multicolumn{1}{l|}{1.293} & 0.037 & -0.04 & -0.106 \\
\ce{Ag2Se} & 0.0 & -0.090 & 0.014 & \multicolumn{1}{l|}{0.153} & 1.734 & 1.740 & 1.890 \\
\ce{Ag2SrO8} & 0.042 & 1.430 & 1.353 & \multicolumn{1}{l|}{1.302} & 0.030 & -0.047 & -0.097 \\
\ce{Ag2Te} & 0.0 & -0.068 & -0.068 & \multicolumn{1}{l|}{-0.068} & 2.281 & 2.289 & 2.282 \\
\ce{AgCrO4} & 0.0 & 0.635 & 0.620 & \multicolumn{1}{l|}{0.621} & -0.060 & -0.060 & -0.060 \\
\ce{AgFe2O4} & 0.0 & 0.472 & 0.235 & \multicolumn{1}{l|}{0.215} & 0.031 & -0.104 & -0.098 \\
\ce{AgNi2O4} & 0.0 & 0.699 & 0.462 & \multicolumn{1}{l|}{0.463} & 0.126 & -0.111 & -0.145 \\
\ce{AgPd2O4} & 0.029 & 0.778 & 0.777 & \multicolumn{1}{l|}{0.778} & -0.087 & -0.088 & -0.087 \\
\ce{As2HO6P} & 0.0 & -0.071 & -0.056 & \multicolumn{1}{l|}{0.035} & 0.284 & 0.313 & 0.446 \\
\ce{As2O3} & 0.0 & -0.078 & -0.079 & \multicolumn{1}{l|}{-0.078} & 0.712 & 0.714 & 0.796 \\
\ce{As2Se3} & 0.0 & -0.064 & 0.124 & \multicolumn{1}{l|}{0.373} & 4.076 & 4.139 & 4.709 \\
\ce{As2Se3} & 0.004 & -0.060 & 0.128 & \multicolumn{1}{l|}{0.376} & 4.080 & 4.143 & 4.713 \\
\ce{AuCrO4} & 0.003 & 0.731 & 0.716 & \multicolumn{1}{l|}{0.717} & -0.061 & -0.061 & 0.013 \\
\ce{AuFe2O4} & 0.024 & 0.511 & 0.274 & \multicolumn{1}{l|}{0.254} & 0.066 & -0.069 & -0.063 \\
\ce{AuNi2O4} & 0.006 & 0.723 & 0.485 & \multicolumn{1}{l|}{0.486} & 0.146 & -0.091 & -0.125 \\
\ce{AuPd2O4} & 0.038 & 0.819 & 0.817 & \multicolumn{1}{l|}{0.818} & -0.063 & -0.064 & -0.063 \\
\ce{AuSe} & 0.0 & -0.123 & 0.033 & \multicolumn{1}{l|}{0.241} & 2.072 & 2.081 & 2.382 \\
\ce{BaH4O3} & 0.027 & 0.139 & 0.036 & \multicolumn{1}{l|}{-0.058} & 0.139 & 0.036 & -0.058 \\
\ce{Bi2O3} & 0.034 & -0.064 & -0.066 & \multicolumn{1}{l|}{-0.064} & 0.155 & 0.156 & 0.155 \\
\ce{Bi2Rh} & 0.023 & -0.061 & -0.058 & \multicolumn{1}{l|}{-0.06} & 3.803 & 3.812 & 3.804 \\
\ce{Bi2Se3} & 0.0 & -0.158 & -0.032 & \multicolumn{1}{l|}{0.146} & 3.314 & 3.357 & 3.782 \\
\ce{Bi2SeTe2} & 0.000 & -0.068 & -0.026 & \multicolumn{1}{l|}{0.034} & 4.135 & 4.163 & 4.319 \\
\ce{Bi2Te3} & 0.000 & -0.075 & -0.075 & \multicolumn{1}{l|}{-0.075} & 4.609 & 4.620 & 4.610 \\
\ce{BiHO6Se2} & 0.0 & 0.301 & 0.315 & \multicolumn{1}{l|}{0.376} & -0.059 & -0.046 & 0.103 \\
\ce{CaH2O2} & 0.014 & 0.217 & 0.051 & \multicolumn{1}{l|}{-0.101} & 0.217 & 0.051 & -0.101 \\
\ce{CaO6S} & 0.041 & 0.642 & 0.536 & \multicolumn{1}{l|}{0.566} & -0.098 & -0.099 & -0.098 \\
\ce{CdH2O2} & 0.024 & 0.088 & -0.078 & \multicolumn{1}{l|}{-0.11} & 0.088 & -0.078 & -0.110 \\
\ce{CoRe2O8} & 0.0 & 0.029 & -0.043 & \multicolumn{1}{l|}{-0.043} & -0.164 & -0.165 & -0.164 \\
\ce{Cr2O5} & 0.0 & 0.447 & 0.422 & \multicolumn{1}{l|}{0.422} & -0.110 & -0.110 & 0.016 \\
\ce{CrCuO4} & 0.024 & 0.496 & 0.481 & \multicolumn{1}{l|}{0.482} & -0.087 & -0.172 & -0.098 \\
\ce{CrO5P} & 0.0 & 0.222 & 0.237 & \multicolumn{1}{l|}{0.368} & -0.070 & -0.029 & 0.165 \\
\ce{CsCHO3} & 0.0 & 0.086 & 0.017 & \multicolumn{1}{l|}{-0.052} & 0.086 & 0.017 & 0.007 \\
\ce{CuFe2O4} & 0.0 & 0.414 & 0.176 & \multicolumn{1}{l|}{0.156} & 0.147 & -0.060 & -0.054 \\
\ce{CuH2O5Se} & 0.004 & 0.438 & 0.445 & \multicolumn{1}{l|}{0.487} & -0.061 & -0.061 & -0.047 \\
\ce{CuH2O5S} & 0.000 & 0.147 & 0.146 & \multicolumn{1}{l|}{0.193} & -0.071 & -0.071 & -0.029 \\
\ce{CuNi2O4} & 0.000 & 0.647 & 0.410 & \multicolumn{1}{l|}{0.41} & 0.249 & -0.060 & -0.094 \\
			\hline
		\end{tabular}
    \vspace{0.5cm}
    \caption{
        List of 2D materials found as highly stable ($\Delta G_{\mathrm{pbx}}\leq -50$ meV/atom) under the CPD analysis in at least one condition of pH and potential relevant to the HER and OER reactions.
    }
    \label{T1}
\end{table}

\clearpage
\begin{table}
    \centering
    
    \vspace{0.5cm}
    \caption{Continuation of Table \ref{T1}}
\end{table}

\clearpage
\section*{CPD of \ce{Ti2C}}
\begin{figure}[ht!]
    \centering
    \includegraphics[width=0.6\textwidth]{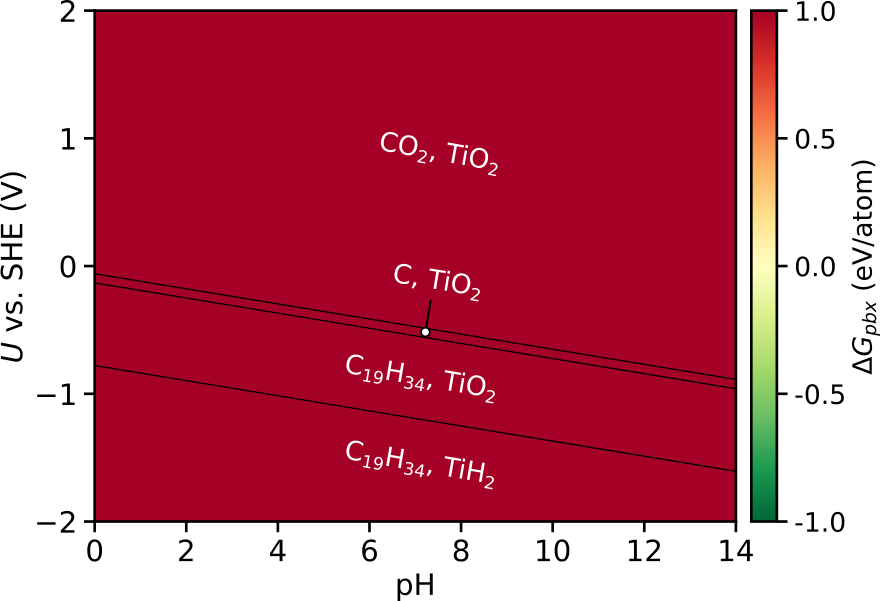}
    \caption{Conventional Pourbaix diagram (CPD) of \ce{Ti2C} against the standard hydrogen electrode (SHE).}
    \label{fig:Ti2C-CPD}
\end{figure}

\section*{Reference electrode potentials}

\begin{figure}[h!]
    \centering
    \includegraphics[width=0.65\textwidth]{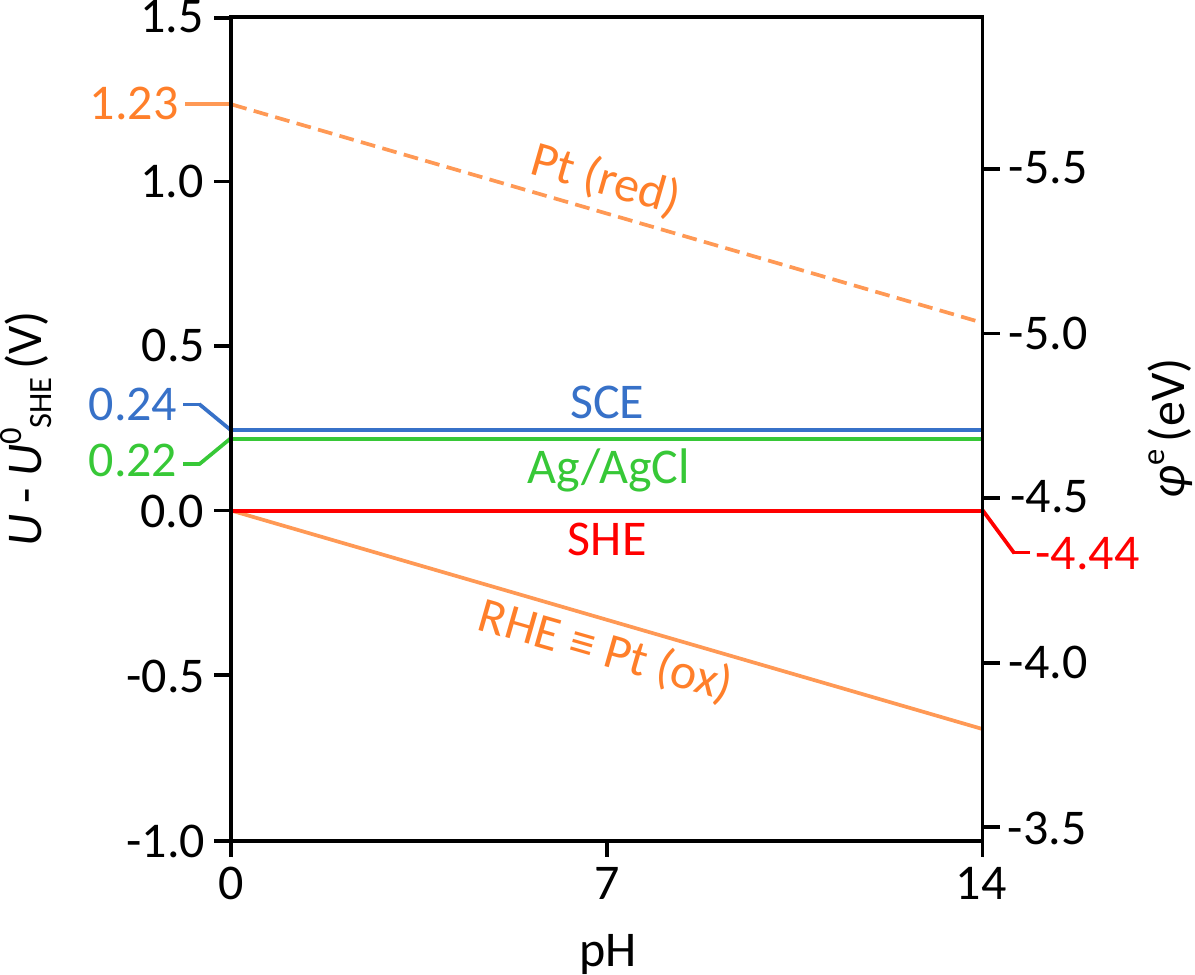}
    \caption{Profile of the potential as a function of the pH of the counter electrodes implemented in both the CPD and SPD framework. The left vertical axis (electrochemical scale) represents the electrode potential relative to the standard potential of the SHE electrode . The right vertical axis (physical scale) represents the potential energy $\varphi ^e$ of electrons relative to vacuum, with $\varphi ^e$=\textit{U}(eV) - 4.44 eV. The potential of the Pt electrode follows the continuous orange line when used against oxidation processes (electrons driven \textit{towards} the Pt electrode). Against reduction processes (electrons drawn \textit{from} the Pt electrode), it follows the dashed orange line.
    }
    \label{fig:electrodepotentials}
\end{figure}

\pagebreak
\section*{List of electrochemical processes modeled in the SPD}
We provide in Fig. S1 a schematic representation of all the early degradation steps(EDS) modeled using the surface Pourbaix diagrams (SPD) framework, subdivided by type. We distinguish between three types of reactions:

\begin{itemize}

    \item \textbf{Semi-catalytic electrochemical EDS}: used for assessing the stability of materials in electrochemical setups, \textit{i.e.} when connected to a reference electrode through an external circuit and under applied potentials. These EDS are labeled as semi-catalytic because only one semi-reaction involves the target material directly, causing its dissolution or passivation. The other semi-reaction takes place at the counter electrode catalytically, i.e. without directly affecting the structure of the electrode material. These EDS are considered when generating potential-pH diagrams.
    
    \item \textbf{Semi-catalytic spontaneous EDS}: used for evaluating the acqueous stability of materials towards spontaneous charge-transfer reactions. bot semi-reaction takes place at the target material surface, although only one involves the surface structure directly.
    
    \item \textbf{Mixed spontaneous EDS}: included as well in the acqueous stability analysis. In this case, both semi-reactions affect the surface structure of the target material.

\end{itemize}

\begin{figure}[ht!]
        \includegraphics[width=0.7\textwidth]{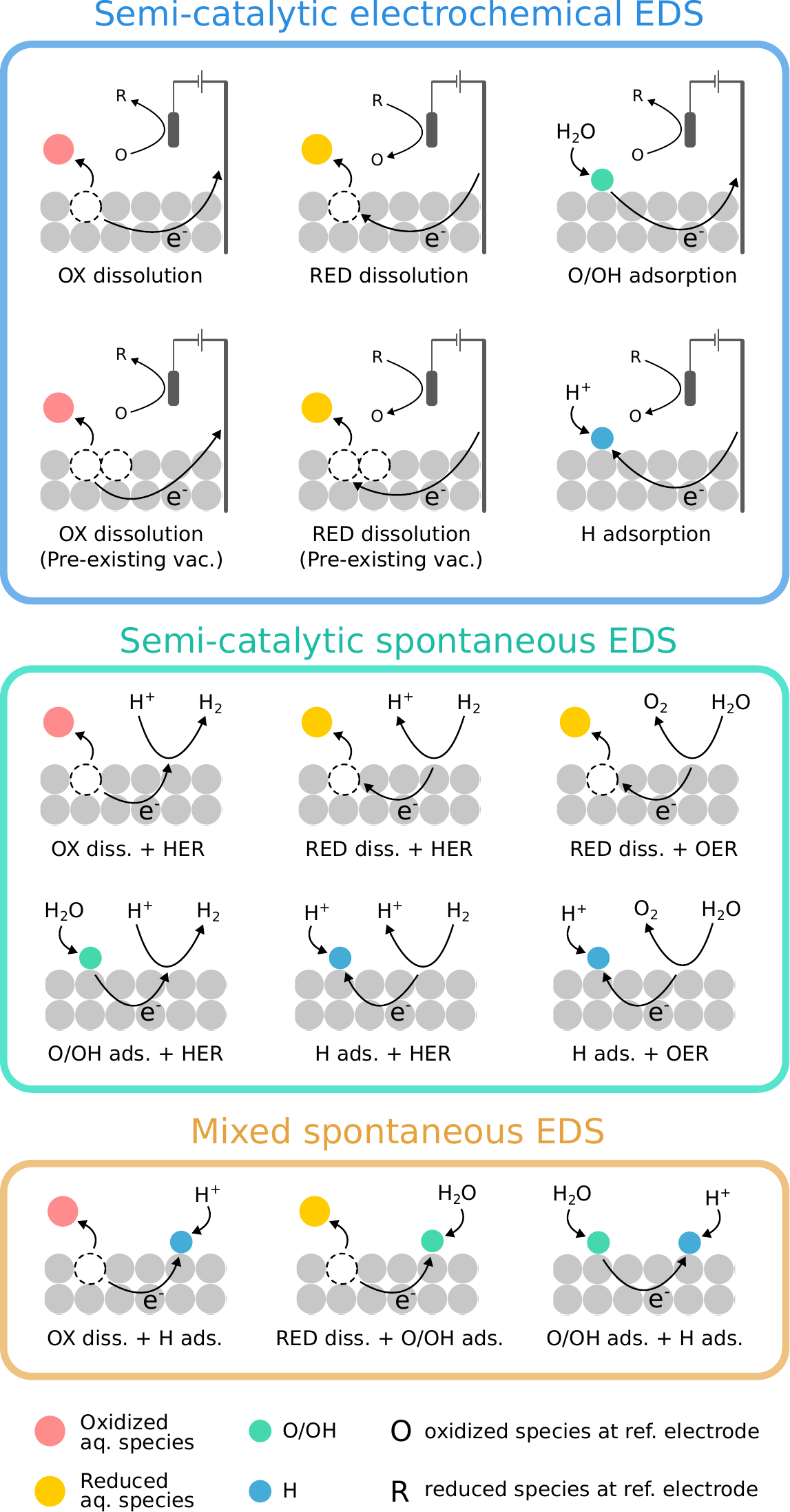}
        \centering
        \caption{
            Illustration of all the EDS included in the SPD analysis.
        }
\end{figure}

\clearpage
\section*{Surface excess correction scheme}

\begin{figure}
    \centering
    \includegraphics[width=\textwidth]{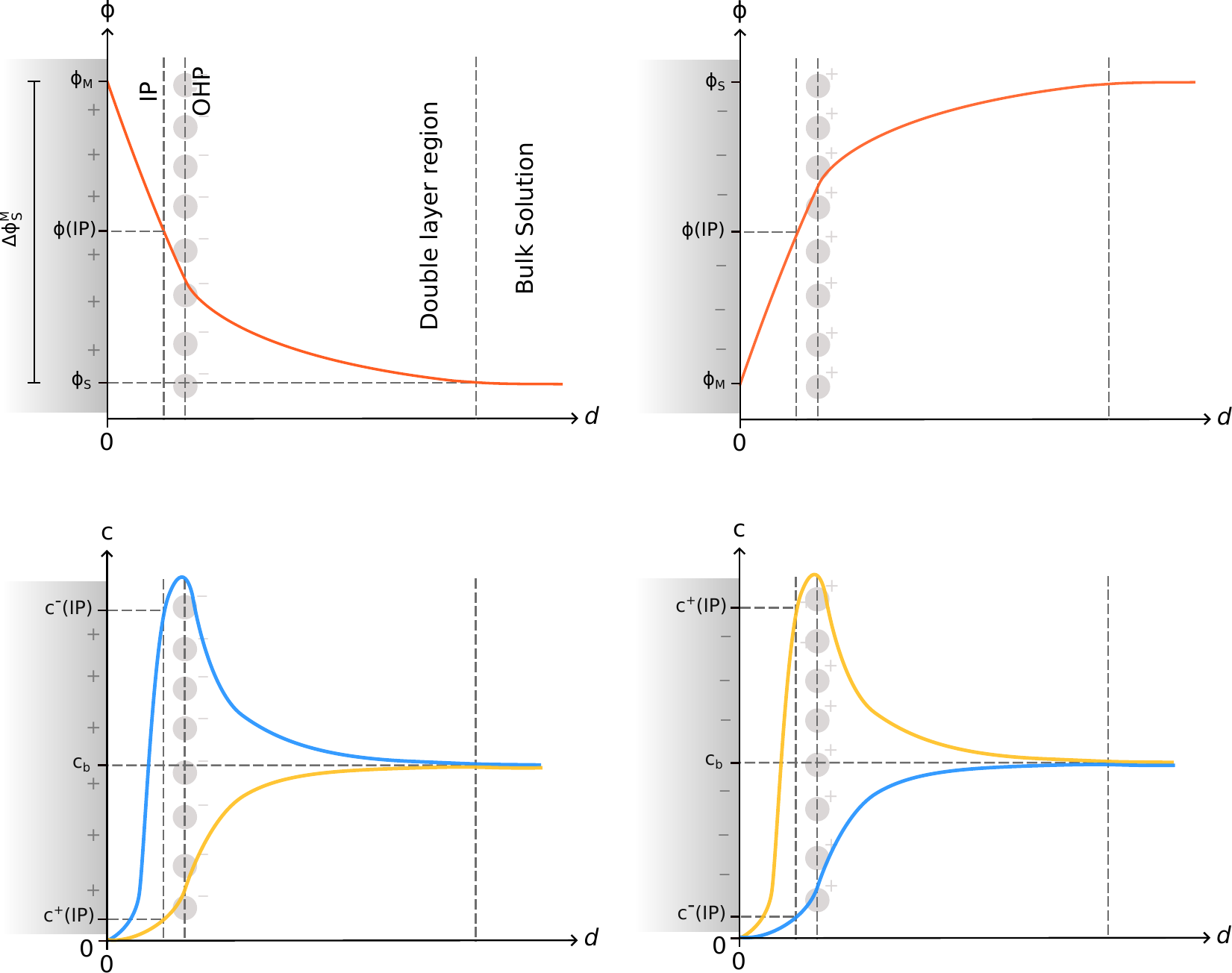}
    \caption{Top panels: qualitative behaviour of the electric potential $\phi$ as a function of the distance $d$ from the surface of a positively charged (left) and negatively charged(right) electrode. $\phi(IP) = (\phi_M + \phi_S)/2$ is the potential at the interaction plane (IP).
    Bottom panels: qualitative behaviour of the concentration of positive (yellow) and negative (blue) ions in the double layer region. $c^{+}(IP)$ denotes the concentration of a positive ion at the IP, while $c^{-}(IP)$ is the equivalent for a negative ion. $c_b$ is the bulk concentration.}
    \label{fig:OHP}
\end{figure}

Below we describe the theoretical background behind the correction scheme implemented in the SPD in order to capture the excess and depletion of charged species at the electrode-solution interface.

The chemical potential of an ion in the solution phase is given by:
\begin{equation}
    \mu = \mu^0 + k_BT \ln c
\label{eq:chempot}
\end{equation}

\noindent where $c$ is the activity of the ion, which we approximate to its concentration.
By using the Boltzmann distribution law, the concentration of an ion at a given point $r$ in space, $c(r)$ can be related to the one in the bulk phase $c^b$ through the expression

\begin{equation}
    c(r) = c^b e^{-W/k_BT}
\label{eq:surf-excess}
\end{equation}

\noindent where $W$ is the work necessary for moving the ion from the bulk solution to point $r$. In absence of short-range ion-ion interactions, this work is purely electrostatic:

\begin{equation}
    W = ze(\phi^S(r) - \phi^S_b)
\label{eq:electric-work}
\end{equation}

\noindent where $z$ is the charge of the ion, $\phi^S(r)$ is the electrostatic potential in the solution at point $r$ and $\phi^S_b$ is the electrostatic potential in the bulk solution.
If not specifically adsorbed on the surface, the ions that sit the closest to the surface are found at an imaginary plane parallel to the electrode surface, called the \textit{outer Helmholtz plane} (OHP). 
As illustrated in Fig. \ref{fig:OHP-IP}, a large part of the overall metal-solution potential difference $\Delta^M_S \phi$ builds up in the thin solution layer delimited by the electrode surface and the OHP.
The distance of the OHP from the electrode surface cannot be determined accurately. In order to obtain a qualitatively correct description, we will assume that the ions able to engage in chemical interactions with the electrode sit at the ``interaction plane" (IP), which pinpoints half of the overall metal-solution potential difference:

\begin{equation}
    \phi(IP)  - \phi^S_b = 1/2\,\Delta^M_S \phi
\end{equation}

\noindent By combining this expression with equation \ref{eq:surf-excess} and \ref{eq:electric-work}, the ion concentration of ion $i$ at the IP is given by

\begin{equation}
    c(IP) = c^b \exp(-\frac{ze\Delta^M_S\phi}{2k_BT})
\label{eq:surf-conc-1}
\end{equation}

\noindent According to Trasatti\cite{trasatti1986absolute}, the absolute potential of an electrode $M$ can be expressed as

\begin{equation}
    \phi^{abs}_M = \Phi_M + \Delta^M_S \phi
\label{eq:absolute-trasatti}
\end{equation}

\noindent where $\Phi_M$ is the work function of the electrode material(in volt). Again from Trasatti, the absolute potential of the standard hydrogen electrode is $\phi^{abs}_{SHE} = 4.44$ V, which can be used in order to obtain the absolute potential of any other electrode through:

\begin{equation}
    \phi^{abs}_M = U + \phi^{abs}_{SHE}
\label{eq:absolute-SHE}
\end{equation}

\noindent where $U$ is the applied cell potential referred to the SHE. By combining eq. \ref{eq:absolute-trasatti} and \ref{eq:absolute-SHE} the metal-solution potential difference can finally be expressed in terms of known quantities:

\begin{equation}
    \Delta^M_S \phi = U + \phi^{abs}_{SHE} - \Phi_M
\label{eq:absolute}
\end{equation}

\noindent and eq. \ref{eq:surf-conc-1} becomes

\begin{equation}
    c(IP) = c^b \exp \left[ -\frac{ze}{2k_BT} \left( U + \phi^{abs}_{SHE} - \Phi_M \right) \right]
\label{eq:surf-conc-2}
\end{equation}

\noindent Eq. \ref{eq:surf-conc-2} can be introduced in the expression of the chemical potential (eq. \ref{eq:chempot}):

\begin{equation}
    \mu = \mu^0 + k_BT \ln c^b - \frac{z}{2} \left( U + \phi^{abs}_{SHE} - \Phi_M \right)    
\label{eq:mu-with-conc-corr}
\end{equation}

\noindent As expected, since the electrode potential affects the surface ions concentration, which in turn affects the reaction energetics, the chemical potential now includes a potential-dependent term.

Note that, if an ionic species is not present in solution prior to the material dissolution producing the ion itself, its bulk concentration $c^b$ is entirely determined to the number of vacancies generated by the process. A realistic value can be obtained by combining quantities relative to the simulated structure with parameters relative to the geometry of the real electrochemical cell:

\begin{equation}
    c^b \simeq \frac{A_{el}}{A_{sc}N_{av} V_{sol}}
\label{eq:bulk-conc-vac}
\end{equation}

\noindent Where $A_{el}$ is the geometrical area of the electrode, $A_{sc}$ is the area of the simulation cell hosting a single vacancy, $N_{av}$ is the Avogadro constant and $V_{sol}$ is the volume of the solution. For example, a 1 cm$^2$ Pt electrode with a vacancy every four unit cells immersed in 1L of solution produces a bulk ion concentration of approximately 2.6$\,\cdot$10$^{-10}$ mol L$^{-1}$. Let us consider consider a Pt surface. The work function of Pt is 5.65 eV. At an applied potential of 1V, the concentration of ions with $z_i=-2$ and $z_i=+2$ at the surface of this electrode (as calculated from eq. \ref{eq:surf-conc-2}) is 5.8 $\,\cdot$10$^{-14}$ mol L$^{-1}$ and 1.15 $\,\cdot$10$^{-6}$ mol L$^{-1}$, respectively. This translates into a difference of about 0.4 eV between their concentration contributions to the reaction free energy. It is clear how the surface excess can depend dramatically on the relative charge state of ion and surface. Assuming one arbitrary concentration for all ionic species may result in significant underestimation or overestimation of their chemical potential and, thus of the overall reaction energetics.

Figure S\ref{fig:OHP} shows the qualitative profile of the electrostatic potential and concentration in the double layer region for electrode surfaces in different charge states.